\documentclass[nopacs,twocolumn,amsmath,amssymb]{revtex4}

\usepackage[OT2,OT1]{fontenc}
\newcommand\cyr{%
\renewcommand\rmdefault{wncyr}%
\renewcommand\sfdefault{wncyss}%
\renewcommand\encodingdefault{OT2}%
\normalfont
\selectfont}
\DeclareTextFontCommand{\textcyr}{\cyr}

\usepackage{graphicx}
\usepackage{amssymb}
\usepackage{mathpazo}
\usepackage{epstopdf}
\usepackage{float}      
\newcommand{\ignore}[1]{}
\newcommand{\be}{\begin{equation}} \newcommand{\ee}{\end{equation}}

\def\ba#1\ea{\begin{align}#1\end{align}}
\newcommand{\bit}{\begin{itemize}}
\newcommand{\eit}{\end{itemize}}

\newcommand{\nn}{\nonumber}  \newcommand{\ra}{\rightarrow}

\renewcommand{\a}{\alpha} \renewcommand{\b}{\beta}

\newcommand{\p}{\partial}

\def\slashb#1{\setbox0=\hbox{$#1$}#1\hskip-\wd0\dimen0=5pt\advance
        \dimen0 by-\ht0\advance\dimen0 by\dp0\lower0.5\dimen0\hbox
          to\wd0{\hss\sl/\/\hss}}

\input{epsf}

\begin{document}

\title{ The Need to Fairly Confront Spin-1 \\
for the New Higgs-like Particle   } 

\author{John P. Ralston} 
\affiliation{Department of Physics \& Astronomy,\\ University of Kansas, Lawrence KS 66045}

\begin{abstract}

Spin-1 was ruled out early in LHC reports of a new particle with mass near 125 GeV. Actually the spin-1 possibility was dismissed on false premises, and remains open. Model-independent classification based on Lorentz invariance permits nearly two dozen independent amplitudes for spin-1 to two vector particles, of which two remain with on-shell photons. The Landau-Yang theorems are inadequate to eliminate spin-1. Theoretical prejudice to close the gaps is unreliable, and a fair consideration based on experiment is needed. A spin-1 field can produce the resonance structure observed in invariant mass distributions, and also produce the same angular distribution of photons and $ZZ$ decays as spin-0. However spin-0 cannot produce the variety of distributions made by spin-1. The Higgs-like pattern of decay also cannot rule out spin-1 without more analysis. Upcoming data will add information, which should be analyzed giving spin-1 full and unbiased consideration that has not appeared before.  

\end{abstract}

\maketitle

Recently the ATLAS\cite{atlas} and CMS\cite{cms} experiments reported a new particle with a mass near 125 GeV. Soon after the Fermilab CDF and D$0$ experiments also reported signals.\cite{fermilabhiggs} The mass value and coupling to several channels make it a prime candidate for the Standard Model Higgs particle. The new particle's spin has not been directly measured. A fair consideration of the spin-1 possibility is needed. The experimental claims ruling out spin-1 are indirect, and based on faulty premises. Analysis is not kinematic, and any firm conclusion that spin-1 can be eliminated either requires new data, new analysis, or the application of theoretical prejudice. 
 
Much rests on observing a bump near 125 GeV in the two photon invariant mass distribution. Consider $s$-channel annihilation proceeding through a vector $ Z'(Q) \ra \gamma (k_{1})+\gamma(k_{2}). $ Let $\epsilon_{Z}^{\rho}$ be the initial state polarization, and $\epsilon_{1}^{\mu}, \, \epsilon_{2}^{\nu}$ the final state photon polarizations. The final state must be symmetric under interchanging all labels of the two bosons. A general 3-vector vertex $V^{\rho \mu \nu}(Q, \ell)$ must have the symmetry \ba V^{\rho \mu \nu}(Q,  \, \ell)=V^{\rho \nu \mu}(Q, \, -\ell); \nn \\ Q=k_{1}+k_{2} ; \:\;\:\: \ell =k_{1}-k_{2}. \nn \ea Assuming Lorentz invariance, and up to terms from gauge-fixing, the most general possibility is \ba V^{\rho \nu \mu}&={1\over \mu^{2}}\sum_{j} \, V_{+j}^{\rho}T_{+j}^{\mu  \nu}+V_{-}^{\rho} (Q^{\mu} \ell^{\nu}+Q^{\nu} \ell^{\mu})+Q^{2}\varepsilon^{\rho \mu \nu \sigma}V_{-\sigma} \nn \\ & +Q^{2}(V_{+}^{\mu}g^{\rho \nu}+V_{+}^{\nu}g^{\rho\mu}) +Q^{2}( V_{-}^{\mu}g^{\rho \nu}-V_{-}^{\nu}g^{\rho\mu} ) \nn \\ & + V_{+}^{\rho}\varepsilon^{\mu \nu \lambda \sigma}Q_{\lambda}\ell_{\sigma}+( V_{+}^{\mu}\varepsilon^{\rho \nu \lambda \sigma}Q_{\lambda}\ell_{\sigma}- V_{+}^{\nu}\varepsilon^{\rho \mu \lambda \sigma}Q_{\lambda}\ell_{\sigma}) \nn \\  & +( V_{-}^{\mu}\varepsilon^{\rho \nu \lambda \sigma}Q_{\lambda}\ell_{\sigma}+ V_{-}^{\nu}\varepsilon^{\rho \mu \lambda \sigma}Q_{\lambda}\ell_{\sigma})
; \\ T_{+}^{\mu  \nu} &=(Q^{2}g^{\mu  \nu}, \,   Q^{\mu} Q^{\nu},\, \ell^{\mu}\ell^{\nu},  \, Q^{\mu} \ell^{\nu}-Q^{\nu} \ell^{\mu}  ); \nn \\ V_{j\pm }^{\rho}& =c_{j\pm}Q^{\rho}+c_{j\mp}\ell^{\rho}. \label{symms}  \ea

Here $c_{j\pm }(k_{1}, \, k_{2}) =\pm c_{j\pm }(k_{2}, \, k_{1}) $ are functions of $k_{1}^{2}, \, k_{2}^{2}, \, Q^{2}$, and $\mu$ is a mass scale to make them dimensionless. The notation implies $c_{\pm}$ in $V_{\pm}$ symbols without index $j$ are independent. Symmetry requires $c_{j-} \sim (k_{1}^{2}-k_{2}^{2})$ times an even function. 

Go to the center of mass frame $Q^{\mu} =(Q, \, \vec 0)$ and $\vec k_{1}=-\vec k_{2}=\vec \ell/2$. On-shell photons imply $Q\cdot \ell=0$, and $Q \cdot \epsilon_{1} =Q \cdot \epsilon_{2} =\ell \cdot \epsilon_{1} =\ell \cdot \epsilon_{2} =0$. Producing the $\gamma\gamma$ signal leaves \ba V^{\rho \nu \mu}=  c_{0+}{Q^{2}\over \mu^{2} }Q^{\rho}g^{\mu \nu} +c_{3+} Q^{\rho}\varepsilon^{ \mu \nu \lambda\sigma}Q_{\lambda}\ell_{\sigma}+{c_{5+} }{Q^{2}\over \mu^{2} }\varepsilon^{\rho \mu \nu \sigma}\ell_{\sigma}. \label{signal} \ea Although $c_{3+}$ can be eliminated with on-shell kinematics it is included for later discussion. Since the interactions are not zero, the spin-1 possibility exists.

More assumptions are needed to limit the possibilities. There are few prohibitions against composite fields. Standard Model particles would need strong new interactions to bind to spin-1 at 125 GeV: 
hydrogenic binding of a top-quark pair needs a coupling constant $\tilde \a \sim 4.5$. That would wreak havoc with known top physics without offering hope for decay patterns close to the Standard Model Higgs. ``Technicolor''-inspired models\cite{techni} assume composite fields couple like the Standard Model Higgs. New experimental limits on techni-spin-1 masses exist\cite{techni}; it is not clear whether they apply.

It is usually assumed that a massive spin-1 theory requires gauge invariance, for which a $Z\gamma \gamma$ vertex points to a new $U(1)$ symmetry. Let $Z_{\mu \nu} = \p_{\mu }Z'_{\nu} -\p_{\nu}Z'_{\mu}$, and $F_{\mu \nu}$ be the corresponding field strength made from the photon field $A_{\mu}$. Under the $SU_{L}(2) \times SU_{R}(2)$ decomposition of the Lorentz group these tensors transform like spin (1,0)+(0,1). The dual $\tilde Z_{\mu \nu}=\varepsilon_{\mu \nu \a \b}Z^{\a \b}$ transforms like (1,0)-(0,1). To make a scalar from three field strengths we need a (0,0) from $((1,0)+(0,1))_{a} \otimes ((1,0)+(0,1))_{b} \otimes ((1,0)+(0,1))_{c}$. Our particle labels are now $abc$. Use $1_{a}\otimes 1_{b} =2_{ab}+1_{ab}+0_{ab}$. Making an invariant with $1_{c}$ needs a spin-1 from $a \otimes b$, but the $1_{ab}$ is antisymmetric in $ab$. The details are interesting. For example, with 12 symmetrization we find: \ba Z_{\mu \nu}'F^{\mu \rho}\tilde F_{\rho}^{\nu} =&{ Q \cdot \epsilon_{Z} \over 2}  \varepsilon_{\rho \nu \a \b}Q^{\nu} \ell^{\a}\epsilon_{1}^{\rho}\epsilon_{2}^{\b}  -{Q^{2} \over 2} \varepsilon_{\rho \nu \a \b}\epsilon_{Z}^{\nu} \ell^{\a}\epsilon_{1}^{\rho}\epsilon_{2}^{\b}. 
\label{ito} \ea When $\epsilon_{Z}^{\nu} \sim Q^{\nu}$ the two terms cancel. When  $\epsilon_{Z}^{\nu}$ is orthogonal to $Q^{\nu}$ both terms give zero; hence the expression is zero due to gauge invariance.

This suggests a ``theorem'' that the $Z'$ sector cannot be gauge invariant\cite{ward} and go to $\gamma \gamma$. As a loophole, recall that the difference of two $U(1)$ gauge fields is gauge invariant: $Z_{-}^{\mu}=Z_{1}^{\mu}-Z_{2}^{\mu}\ra Z_{1}^{\mu}-\p^{\mu}\theta -Z_{2}^{\mu}+\p^{\mu}\theta =Z_{-}^{\mu}$. Actually any linear combination of two fields is invariant under a corresponding subgroup of local $U_{1}(1)\otimes U_{2}(1)$. That allows a gauge invariant mass term to exist. But the attractive features of perturbative unitarity and renormalizability of gauge theories stem from being able to use a propagator orthogonal to $Q^{\rho}$, which would decouple from the vertex making real $\gamma \gamma$. 

Regardless of theoretical models we don't see a reason for experiments to assume gauge invariance for new physics. Two experimentally allowed interaction Lagrangians are \ba  L=a_{1} \p \cdot Z' F \cdot F +a_{2} \p \cdot  Z' F \cdot \tilde F.  \label{way1}\ea These come from the symmetric spin-0 products. The spin-2 modes from $Z'$ are found in $\xi_{\mu \nu} =  \p_{\mu} Z_{\nu}'+\p_{\nu}Z_{\mu}'$. Using them produces two possibilities \ba L  =b_{1}\xi_{\mu \nu}F^{\mu \rho}F_{\rho} ^{\nu}+b_{2}\xi_{\mu \nu}F^{\mu \rho}\tilde F_{\rho} ^{\nu}. \label{way2} \ea Algebra with on-shell kinematics shows all the above reduce to combinations of terms listed in Eq. \ref{signal}. 

A constraint $\p \cdot Z'=0$ can be introduced to define Eq. \ref{signal} to be zero. But any interaction can be defined to be zero. The constraint is related to theoretical desires for relativistic ``spin-1'' to have literally three degrees of freedom. A related dodge introduces a scalar field $\phi$ and writes $Z^{' \mu} =Z_{\perp}^{' \mu}+ \p^{\mu} \phi$ where $ \p \cdot Z_{\perp}' \equiv 0$. Using that too literally would wrongly assert a scalar and not a vector field interacts in Eq. \ref{signal}. Actually it proves that {\it If and When} $Z^{'\mu} \sim Q^{\mu}$ it is hard to tell the effects of Eq. \ref{signal} from interaction with a scalar. When the photons are not on-shell, and in the important decay to $ZZ$ the interactions of Eq. \ref{signal} (as well as the remaining terms of Eq. \ref{symms}) cannot be replaced by a scalar field, leading to signals we discuss below.

General principles restrict very little, while more detail comes from models. The Proca theory coupled to sources, which is closely related to Stueckelberg models\cite{stuke}, has a more subtle treatment of $\p \cdot Z'$. The equation of motion is $ \p_{\mu}Z^{'\mu \nu}+m_{Z'}^{2}Z^{'\nu}=j^{\mu}$.  In the free-field approximation that $j^{\mu} \ra 0$ the equation implies $\p \cdot Z' \ra 0.$ Hence asymptotic states have three modes. Analysis also shows the momentum conjugate to $Z^{0}$ does not exist, meaning it is not dynamical, but dependent. Yet under interactions the constraint can be revised, which is tracked by the corresponding propagator $G_{\lambda \rho}$: \ba G_{\lambda \rho}=-i{ g_{\lambda \rho} -Q_{\lambda}Q_{\rho}/m_{Z'}^{2} \over Q^{2}-m_{Z}^{'2}+ i m_{Z'}\Gamma_{Z'}}. \nn \ea This is not orthogonal to $Q^{\rho}$, hence couples to our vertex to produce an interaction.

Thus spin-1 is allowed for the $LHC$ observations, but producing $Z'$ appears to need a non-conserved current in the amplitude considered. Non-conserved currents exist in Nature and the decay $\pi^{0} \ra \gamma \gamma$ occurs because a certain axial current amplitude {\it cannot consistently be conserved} when related vector currents are conserved. That history tied to chiral anomalies suggests an axial spin-1 field. Recent activity\cite{axials} invokes massive axial spin-1 particles to explain the top-quark charge asymmetry observed at Fermilab\cite{top}.

It is widely believed the spin-1 possibility was ruled out early. Actually CDF and D$0$\cite{fermilabhiggs} do not mention the word ``spin.'' ATLAS\cite{atlas} writes that ``The observation in the diphoton channel disfavours the spin-1 hypothesis[140,141].'' CMS\cite{cms} writes in four places that the two-photon decay implies the new particle's ``spin is different from one [129, 130].''  The references cited are the Landau-Yang theorems\cite{landau,yang}. Reviewing the derivation shows why the theorems are inadequate. 
 
Yang's method\cite{yang} lists the joint polarization states of two final state photons in their center of mass frame. With $R$, $L$ representing right and left-handed helicities there are four possible combinations written $\Psi^{RR}$, $\Psi^{RR}$,$\Psi^{RL}$, $\Psi^{LR}$, $\Psi^{LL}$. The state $\Psi^{RL}$ transforms like $e^{2 i\phi}$ when the coordinate system is rotated by angle $\phi$  around the $z$ axis, hence has spin angular momentum $S_{z}=2$. Yang says this is forbidden to come from spin-1. Bose symmetry is used to complete a table of selection rules. Nothing of gauge invariance, Lorentz invariance, or momentum dependence is mentioned. 

CMS \cite{cms} also cites Choi, Miller and Zerwas\cite{choi} on identifying the Higgs spin and parity. The paper's Eq. 7 claims a general amplitude for spin-$J$ to produce back-to-back $Z$ at angles $\Theta, \, \Phi$ takes the form $T_{\lambda_{1} \, \lambda_{2}}d^{J}_{ \, m, \, \lambda_{1}-\lambda_{2}}(\Theta) e^{i(m- \lambda_{1}+\lambda_{2})\Phi}$, where the the reduced vertex $T_{\lambda_{1} \, \lambda_{2}}$ ``depends only on the helicities ($\lambda_{j}$) of the two real $Z$ bosons''. This applies Yang's method to arbitrary spin. Yang's argument is also repeated almost verbatim in the 2008 paper of Keung, Low and Shu\cite{choi}, which reiterate the rotational and parity transformation properties of polarization products for massive final states.

The general method of enumerating amplitudes by polarizations alone produce a few paradoxes. For example, it implies that spin-1 cannot decay to two spin-0 and conserve angular momentum, so $\vec \rho_{770}(k_{3}) \ra  \pi(k_{1})\pi(k_{2})$ should be impossible. The flaw is revealed in the vector current vertex $k_{1}^{\rho}-k_{2}^{\rho}$ which accounts for the {\it orbital} angular momentum neglected in the method of counting helicities. Landau's paper\cite{landau} classifying the transformation properties of two-photon wave functions avoids the mistake. For example Landau includes the spin (``\textcyr{spin}'')-2 from polarization before dealing with the orbital (``\textcyr{orbital\cyrsftsn nym}'') angular momentum needed to make total $\vec J = \vec L+\vec S$. Our Eq. \ref{way1} is one of Landau's processes classified under total angular momentum $J=0$, parity =$\pm1$. Landau's discussion does not extend to Lorentz invariance or virtual particles, and applying it to decays will cause an error unless all modes are considered. 

Giving spin-1 fair treatment needs experiments to consider the signals. In producing $\gamma \gamma$ most of the amplitude will come from the region dictated by the width, $Q^{2}-m_{Z}^{'2} \sim  m_{Z'}\Gamma_{Z'}$. In this region the phase of the propagator varies rapidly. The phase will be important in interference. This matters because Standard Model amplitudes to continuum $\gamma \gamma$ states are large compared to most new-physics amplitudes. In generating the cross section, the interference of a relatively large amplitude with a small new-physics amplitude may be much larger than the new amplitude squared. Thus seeking interference can be an effective way to find new physics.

The cross section $d\sigma/dQ^{2}$ to measure $\gamma \gamma$ with invariant mass-squared $Q^{2}$ is the sum of the squares of many amplitudes. The calculation summing over final $\gamma\gamma$ polarizations can be written \ba {d\sigma \over dQ^{2}}  \sim \sum_{X \mu \nu}(  M_{X}^{\mu \nu}+  M_{X \rho} & V^{\rho \mu \nu } ) \nn \\ & \times (  M_{X }^{\mu \nu ^{*}}+ M_{X \rho'}V^{\rho' \mu \nu }) . \nn \ea Here $\sum_{X} \, M_{X \rho}M_{X \rho'}^{*}$ is the density matrix to produce $Z'$ from the initial state, summed over final states $X$, and including phase space factors and the $Z'$ propagators. This would be calculated using the parton model and $Z'$ production channels.

Attaching the propagator to one vertex gives \ba G_{\lambda \rho}V^{\rho \mu \nu} =i c_{5+}{ Q^{2}-m_{Z'}^{2} \over m_{Z'}^{2}  (Q^{2}-m_{Z}^{'2}+ i m_{Z'}\Gamma_{Z'}) } {Q_{\lambda}Q_{\rho} \over Q^{4}} \epsilon^{\rho \mu \nu \sigma}\ell_{\sigma}. \label{dip} \ea We use the identity only for the $V^{\rho} \sim Q^{\rho}$ term we need. Recall from Eqs. \ref{way1}, \ref{way2} the amplitudes may scale like $Q^{2}/\mu^{2}$, or higher powers of $Q$. Taking into account the dimensions and scales shown, a schematic calculation takes the form \ba { d\sigma\over dQ^{2} }\sim \sum_{j} |   {  \hat M e^{i \phi} \over Q^{2}}  + i \tilde c_{j}{ Q^{2}-m_{Z'}^{2} \over m_{Z'}^{2}  (Q^{2}-m_{Z}^{'2}+ i m_{Z'}\Gamma_{Z'}) } |^{2} +{ d\sigma_{inco}\over dQ^{2} }. \label{model} \ea Here $ \hat M$ is a proxy for interfering amplitudes scaled to be dimensionless, and absorbing the overall scale, $\tilde c_{j}$ stand for the production and decay factors lumped together. Symbol $d\sigma_{inco}/ dQ^{2} $ represents channels added incoherently, and expected to dominate the background.

\begin{figure}[htbn]
\begin{center}
\includegraphics[width=3.5in]{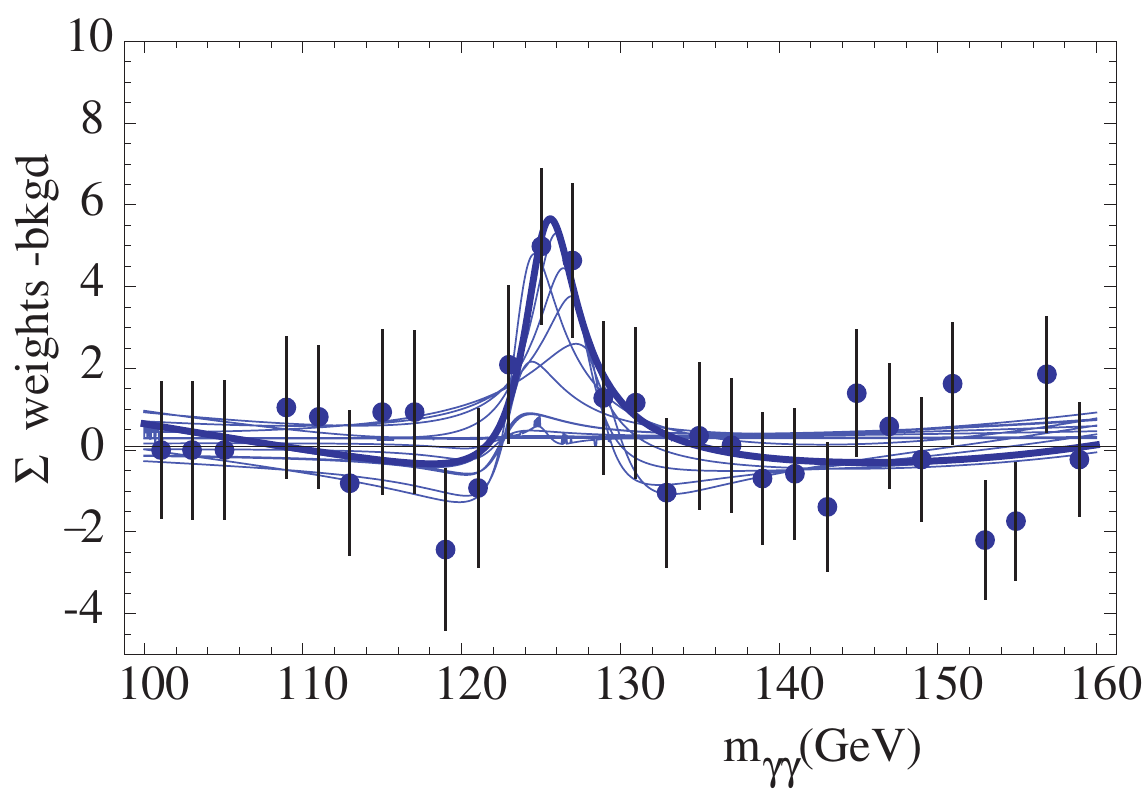}
\caption{ Proof of principle that residuals of the invariant mass distribution $d \sigma / d Q$ versus $Q$ are consistent with a spin-1 resonance. The thick curve (blue online) is the best fit with $m_{Z'}=125.2$ GeV, $\Gamma_{Z'}=4.1$ GeV, phase $\phi=1.33$, and other parameters cited in the text. The background curves are best-fits for the full range $0 \leq \phi < 2 \pi$ in steps of $\pi/12$. They illustrate the variety of dip-bump structure, which may have been seen in both ATLAS and CMS data. Residuals from ATLAS weighted sample. }
\label{fig:AtlasFitVary.eps}
\end{center}
\end{figure}

The $\gamma \gamma$ invariant mass distribution comes from convolution over initial state parton distributions, the model of the interaction, and integration over the undetected final states and acceptance of the detectors. Full consideration of spin-1 would seem to need that level of detail. Fortunately ATLAS has presented the residuals of the data for $d \sigma/dQ$ relative to backgrounds. Figure \ref{fig:AtlasFitVary.eps} shows a simple illustration that a vector particle can fit the residuals well. Using $\hat M=1$, the best fit values are $m_{Z'}=125.2$ GeV, $\Gamma_{Z'}=4.1$ GeV, $\phi=1.33$, and $\tilde c=5 \times 10^{-5}m_{Z'}^{2}/GeV^{2}$; only one $ \tilde c_{j}$ was used. The calculation was adjusted by an overall normalization $\kappa=6.42$, and a constant of 0.16 was subtracted. The value of $\chi^{2}/NF=13.2/24$ is low, mostly due to many points over the reported range $100<Q<160$ GeV that are naturally close to zero. Since the experimental groups will not release even the un-binned data used to make histograms we are forced to digitize the published figures. 

Figure \ref{fig:AtlasFitVary.eps} also shows the effects of different phases $\phi$. The background curves (thin lines) come from fixing $\tilde c$ to the value above and $\phi = n\pi/12$ for integer $n=0-11$. The other parameters were then evaluated at their best-fit values.  Most phases produce a dip-bump structure. We observe that fine structures of two independent experiments do suggest a dip-bump structure has been seen. There are several (at least 5 total) points forming a dip at lower mass than the bump in the same region of {\it both} the ATLAS and CMS data. We suggest the pole region should receive careful scrutiny in future analysis.

The fitted values are tentative but perhaps provide an order of magnitude for the spin-1 possibility. The value of $\tilde c \sim 0.8$ (relative to $\hat M$=1) indicate two comparable interfering suffice to make a bump. Justifying why the coherently interfering parts should be the right size seems arbitrary. Note the fit includes effects of experimental resolution, as the physical width is unknown. Smaller widths make sharper dip-bumps. Due to a lack of symmetry they are not always erased by smearing. Relative phases are generally momentum dependent. Different probes can shift the apparent pole-mass position significantly. Using two amplitudes for $\gamma \gamma$, and all possible for $ZZ$ production allows great complexity of signals. The apparent width and dip-bump structure might also be due to more than one resonance (whether or not spin-1). 

Giving different hypotheses fair treatment will also consider their discriminating signals. As a rule angular distributions are needed to determine spin. The angular distribution of the $\gamma \gamma$ channel will be background-dominated, so that $ZZ \ra 4 \, lepton$ channels with low backgrounds tends to be more sensitive. By Lorentz and gauge invariance the only possible amplitude for $J^{P}=0^{\pm} \ra \gamma \gamma$ or $ZZ$ go like $\epsilon_{1}\cdot \epsilon_{2}$ and $\varepsilon_{abcd}\epsilon_{1}^{a} \epsilon_{2}^{b}k_{1}^{c}k_{2}^{d}$, for $P=\pm$, respectively. The same angular observables are produced by $c_{0+}$ and $c_{3+}$ terms: thus signals ``confirming'' spin-0 do not rule out spin-1. Yet there are spin-1 distributions that can rule out spin-0. Production of a vector particle in unpolarized hadron-hadron collisions generally leads to a weighted mixture of longitudinal and transverse modes. With the $c_{5-}$ term the transition from a transverse $Z'$ to two vector gauge particles needs one to be longitudinal and one transverse, which is impossible with a spin-0 Higgs. That provides an example using $Z'$ polarizations that cannot be mistaken for a scalar field. 

Finally one may examine the assumption of Lorentz invariance. Fundamental Lorentz symmetry violation is of great interest. More generally the collisions have a preferred rest frame ``medium'' that break Lorentz symmetry. Photons have a longitudinal mode in a medium with free charges: this can occur with or without a transition to a new phase of matter. Conversion of a longitudinal mode to an observed transverse mode needs very little interaction. Real photons are not always pristine, pointlike probes. For one thing the photon mixes significantly with the $\rho$ meson, which is a great complication. 

The $LHC$ and Fermilab experiments have discovered a new resonance whose spin is unknown. If the the spin-1 possibility can be ruled out it should not come from methods relying on incomplete enumeration of amplitudes or theoretical prejudice. Instead more experimental data, which is expected very soon, should guide the way. Exploring the possibilities of the unexpected without bias should be welcome, and possibly the best road to finding ``new physics''.

{\bf Acknowledgments:}  Alice Bean, KC Kong, Danny Marfatia, Mat McCaskey, Doug McKay and Graham Wilson generously contributed information and helpful comments. Work supported in part under DOE-HEP grant number DE-FG02-04ER14308.


\begin{thebibliography}{99}
\bibitem{ward} General tests of symmetries will depend on a particular theory and its Ward identities.
\end{thebibliography}
\end{document}